\documentstyle[12pt,cite,epsfig,a4]{article}

\begin{document}

\hfill Nijmegen preprint

\hfill September  1996

\hfill HEN-398

\vspace{0.5cm}

\begin{center}

\bigskip
{\Large\bf Multiparticle Clusters and 
Intermittent Fluctuations}                           
\vspace{1.7cm}

{\large S.V.Chekanov\footnote[1]{On leave from
Institute of Physics,  AS of Belarus,
Skaryna Av.70, Minsk 220072, Belarus},
}

\medskip

{\it High Energy Physics Institute Nijmegen
(HEFIN), University of Nijmegen/NIKHEF,\\
NL-6525 ED Nijmegen, The Netherlands}

\vspace{0.7cm}

{\large V.I.Kuvshinov}

\medskip

{\it Institute of Physics,  AS of Belarus,
Skaryna Av.70, Minsk 220072, Belarus}

\vspace{0.7cm}

Presented at 5th Ann. Int. Seminar
``Non-Linear Phenomena in Complex Systems'' 
Minsk, Belarus, February 1996 

\end{center}

\vspace{1.0cm}

\begin{abstract}
An approach  for 
understanding the behavior of multiplicity distributions
in restricted phase-space intervals derived on the basis of global
observables is proposed. 
We obtain a unifying connection between local multiparticle
clusters  and the 
scale-invariant power-law behavior of normalized factorial
moments. The model  can be used 
to describe  multiparticle processes in terms of  a 
decomposition of the observed intermittent signal into
contributions from clusters with varying number of particles.
\end{abstract}

\bigskip


\newpage
\section{Introduction}
\label{sec:intro}

An inverse power-law dependence of  normalized
factorial moments (NFMs) $F_q(\delta )$ on the size of a phase-space
bin $\delta$ (intermittency phenomenon) obtained from 
multiparticle production  in high-energy experiments can serve as  a
signal for dynamical local fluctuations with self-similar structure \cite{on1}.
For a single bin, this power-law behavior can be written as
\begin{equation}
F_q(\delta )\equiv\frac{\langle n^{[q]}\rangle}{\langle n\rangle^q}\propto
\left(\frac{\Delta}{\delta}\right)^{\phi_q}, \qquad
n^{[q]}=n(n-1)\ldots (n-q+1),
\label{1o}
\end{equation}
where $n$ is the
number of particles in the phase-space bin of size $\delta$ in which
these local multiplicity fluctuations are investigated. The angular
brackets imply an averaging over all events in the sample and
$\Delta$ is a full phase-space interval defined in rapidity,
azimuthal angle, transverse momentum or a combination of these variables.
The intermittency exponents $\phi_q$ are related to the so-called 
anomalous fractal
dimensions $d_q$ and the R\'enyi dimensions $D_q$ as follows
\begin{equation}
d_q=\frac{\phi_q}{q-1},  \qquad D_q=D(1-d_q),
\label{2o}
\end{equation}
where $D$ is the topological dimension of  phase space.  If particles
are randomly distributed  in  phase space, then $\phi_q=0$ and $D_q=D$. For
monofractal multiplicity distributions,  $d_q=const$. 
For a multifractal behavior, 
$d_q$ is a function of the rank $q$ of the NFMs.

The scale-invariant fluctuations manifest themselves as ``spikes'' -
clustering of many particles in small phase-space bins for single
events - and are common to many areas of physics
(see  \cite{on2,on3,on4} for reviews).

The factorial-moment method and the concept of intermittency in 
high-energy physics have been borrowed from the theory of turbulence. 
There,  
continuous probability densities are used 
as the mathematical tool for a theoretical
description of  fluctuations \cite{on1}. The main problem
in such an approach is the comparison with experiments having 
a finite multiplicity 
of particles for single events.
However, if  statistical noise has a Poissonian nature, then the
method of NFMs follows immediately 
due to its suppression of the statistical
fluctuations caused by the finiteness of the 
number of particles in restricted phase-space bins.
Then, the values of the NFMs are equal to the usual 
normalized moments obtained from probability densities. 
That is why, from a  theoretical point of view,
attempts have been made to understand intermittency also via the analysis of
usual moments obtained from models borrowed from hydrodynamics 
\cite{on1,on5,on6,on7}.
However, little attention has so far been 
devoted to a systematic treatment of dynamical
models involving a finite number of particles in phase space.

In this paper, we develop the method for calculation
of the NFMs making use of 
discrete  probability distributions, i.e., {\em a priori}
taking into account
the finiteness of the number of particles in a
phase-space interval. 
Such an approach allows us to derive a multiplicity distribution in small
phase-space bins from {\em global} characteristics of 
samples with correlations between particles.
In this context, this paper may
be considered as a continuation of the study of the projection 
method for the transition in probability distributions from  full 
to restricted phase-space intervals \cite{on8,on9}.

The method is based  on a natural assumption about the existence of
multiparticle clusters in  phase space and,
for a given cluster-size distribution, can lead to an exact solution for
the rise of NFMs with decreasing  $\delta$.
Usually, some degree of arbitrariness exists in the
definition of clusters.
Here, a cluster in an individual event
is considered in a general and traditional
sense - as a bunch of
many particles  with a very small extension in
the  phase space under investigation. Let us note that
such concept of spikes is wider than that of clusters, since 
occurrence of the cluster is caused by dynamical reasons while
the spike may have a purely statistical nature.

\section{General Formalism}  
\label{sec:lfm}

First, let us define notations which will be used throughout this paper.
The generating function $G(z)$  for
a  multiplicity distribution  $P_n$  and the unnormalized factorial
moments $\tilde F_q$ are defined as follows
\begin{equation}
G(z)\equiv\sum_{n=0}^{\infty}P_nz^n ,
\qquad \tilde F_q\equiv \langle n^{[q]}\rangle \equiv G^{(q)}(z)\mid_{z=1}.
\label{3o}
\end{equation}
The normalization condition $\sum_{n=0}^{\infty}P_n=1$ leads to
$G(z=1)=1$.
Then, the NFMs $F_q$ for the multiplicity distribution 
$P_n$ are given by the relation
\begin{equation}
F_q= \frac{\tilde F_q}{\tilde F_1^q}.
\label{3oo}
\end{equation}
A capital letter $N$ will be used to specify the number of particles
in the full phase space of size $\Delta$, 
and $n$  represents the number of particles
in a restricted phase-space interval of size $\delta$, so that
$n\le N$ for  $\delta\le\Delta$.
 
Let us consider a collision between two particles 
yielding exactly $N$
final particles  for each event in some full 
phase space of size $\Delta$ with
a topological dimension $D$. 
Let us divide the full phase-space volume into
$M^D$ non-overlapping bins of size
\begin{equation}
\delta =\frac{\Delta }{M^D}.
\label{4o}
\end{equation}

In this paper for simplicity we will consider the case of a flat
phase-space distribution, i.e., 
each infinitely small cell of  phase space is regarded
equally probable.
In this case, none of the quantities characterizing 
fluctuations depend on the
position of the phase-space bin under study.
To compare the model predictions with
the experimental non-flat distributions, therefore, 
the transformation of
a given non-flat  distribution into a flat one 
should be performed \cite{on10}.

The intermittent
fluctuations in physical systems may manifest 
themselves as localized dynamical spikes
(multiparticle clusters) in individual events.  Let us mention
briefly two well-known extreme  cases of phase-space  distribution:

i) If all $N$ particles are equally distributed in a given  
phase-space volume, then the multiplicity distribution
for particles in $\delta$ is a positive-binomial
one \cite{on12},  
with a generating function $G_N(\delta, z)$ of the form
\begin{equation}
G_N(\delta, z)=(pz + g)^N ,  \qquad g=1-p ,
\label{5o}
\end{equation}
\begin{equation}
p=M^{-D}=\frac{\delta }{\Delta }.
\label{6o}
\end{equation}
From (\ref{3oo}),  we have the following form of the NFMs
\begin{equation}
F_{q}=\frac{N^{[q]}}{N^q},  \qquad q=1, 2,\ldots  ,N.
\label{7o}
\end{equation}
The extreme case thus gives $\phi_q=d_q=0$, $D_q=D$. This means
that all spikes observed in single events have a purely statistical nature.
Note that $F_{q}<1$ for all $q$.  However, for $N\to\infty$,
the distribution (\ref{5o}) tends to a Poissonian one with $F_{q}=1$.

ii) Another (unlikely) situation occurs if all $N$ particles group in one
single point-like cluster for all possible events. Then the generating
function is \cite{on9}
\begin{equation}
G_N(\delta, z)=pz^N + g ,  \qquad g=1-p ,
\label{8o}
\end{equation}
with  $p$ of the form (\ref{6o}). 
This distribution law emphasizes that,
for a given bin size $\delta$, 
only two possibilities can occur: either all
$N$ particles are found in the bin or none.
For the NFMs, one gets
\begin{equation}
F_{q} =
\frac{N^{[q]}}{N^q}\left(\frac{\Delta}{\delta }\right)^{q-1},
\qquad q=1, 2, \ldots  ,N.
\label{9o}
\end{equation}
This is a maximum possible intermittency ($\phi_q=d_q=1$, $D_q=0$).
From a geometric point of view, this  case corresponds to a point-like
object having topological dimension zero.

The two examples presented above 
lead to the two main theoretical 
questions: how do such  dynamical spikes  lead
to the actual form of intermittency (\ref{1o}) with a small 
(but non-zero) exponent $\phi_q$
and how does one  
construct the multiplicity distribution 
for this case. 
The aim of the present analysis is to
derive a probabilistic scheme of the 
realistic intermediate situation with
$0<\phi_q<q-1$ in (\ref{1o}), using 
known global characteristics of the sample.

\medskip
In our approach we restrict ourselves  to the following cases:

1) All single particles (monomers) and
multiparticle clusters are placed in phase space
independently, without any correlations.
Within the analytical model to be discussed below, all
possible correlations are those between particles inside each cluster.

2) We shall consider the simplified case
of treating the clusters as ``point-like objects''.
In fact, we suppose that the
probability that  the cluster is emitted on the boundary of the
restricted  phase-space interval is very small, i.e.,  
$\delta\gg\delta_{min}\sim\Omega$, where
$\Omega$ is the cluster size. Since intermittency is defined via
the scale-invariant ratio $\Delta/\delta =M^D$, we can rewrite this
condition as
\begin{equation}
\frac{\Delta}{\Omega}\sim\frac{\Delta}{\delta_{min}}\gg M^D .
\label{97o}
\end{equation}
Hence, for a given $M$, a cluster can be considered as  
a point-like object if 1) the size $\Omega$ of the cluster is
very small; 2) the total phase-space
volume $\Delta$ of multiparticle  production is large. 
For  real experiments, 
in fact, condition (\ref{97o}) means that we consider the case when the
number of phase-space bins $M$ is not very large.

\medskip
Consider the following structure of a fluctuation.
Let us assume that  not only single particles but also
point-like multiparticle clusters can exist in single events. 
Let each of them contains
exactly $m$ particles ($m$-particle clusters), where $m=2,3,4\ldots $.
Then, only the two cases are possible: a cluster is inside
$\delta$ or outside of this interval. We assume that the 
cluster phase-space distribution
is flat. 
If only multiparticle clusters with
a fixed $m$ exist, then the 
multiplicity distribution
to have $n$ particles (belonging to clusters) inside $\delta$
has the binomial-like form:
\begin{equation}
P_n(m,N_m)=\left\{\begin{array}{ll}
0, & \mbox{if $n\ne m\>s$,\   $s=1,2\ldots$}\\
C^{\frac{n}{m}}_{N_m}p^{\frac{n}{m}}g^{N_m-\frac{n}{m}}, 
& \mbox{otherwise}, 
\end{array}\right.
\label{100o}
\end{equation}
where $n=0,1,2\ldots mN_m$,
$p$ is given by (\ref{6o}) and $N_m$ is the total number of 
$m$-particle clusters in full phase space. 
$C^n_N$ are the binomial coefficients. The $N_m$ is a constant for the
sample of events 
and will be weighted over all possible samples below. The generating function
for (\ref{100o}) is
\begin{equation}
G_{N_m}^{\{m\}}(\delta ,z)=
\sum_{i=0}^{N_m}z^{m\>i}C_{N_m}^ip^ig^{N_m-i}=(pz^m+g)^{N_m}.
\label{10o}
\end{equation}

As mentioned in the introduction, we will consider the case
of no correlations between different clusters  and 
monomers. Then, the generating function 
for the probability of having $n$ particles in
$\delta$, if $m$-particle clusters ($m=2,3\ldots$) and
monomers ($m=1$) exist, is the product of  the generating functions
(\ref{10o})
\begin{equation}
G_R(\delta, z, N)=\prod_{m=1}^{N}G_{N_m}^{\{m\}}(\delta ,z) ,
\label{11o}
\end{equation}
where $G_{N_1}^{\{1\}}$ is the generating function
(\ref{5o}) for  uncorrelated  monomers.
Note that the generating function (\ref{11o}) satisfies the
normalization condition 
$G_R(\delta, z=1, N) = 1$ by construction.
The set $R\equiv \{N_m\}\equiv (N_1, N_2, ..., N_N)$
represents a cluster configuration (cluster-size distribution) which 
satisfies the following  constraint
\begin{equation}
N=\sum_{m=1}^{N}m\,N_m .
\label{12o}
\end{equation}
Note that the extreme case $N_N=1$, when all $N$ particles form a
single cluster-monomer, is possible only if $N_1=N_2=\ldots N_{N-1}=0$.

A real situation is expected 
to be more complicated, because, for a fixed $N$, the
events can differ one from another by the cluster configuration
$R$. 
In this case,  $R$ is a random variable. Then, one 
averages the $G_N(\delta ,z)$ 
over all possible configurations, i.e.,
\begin{equation}
\bar G(\delta ,z, N)=
\sum_{R=1}^{\infty}P_R^{\mathrm{conf}}
G_R(\delta, z, N)=
\left<G_R(\delta, z, N)\right>_R ,
\label{13o}
\end{equation}
where $P^{\mathrm{conf}}_R$ is  a probability distribution depending
on $N$ and controlling the relative weight  of the multiplicity 
distributions  for different cluster
configurations.

Moreover, for actual multihadronic systems produced
in high-energy experiments, one needs to average over all events
with different $N$. Then,
\begin{equation}
G(\delta ,z)=\sum^{\infty}_{N=0}P_N\bar G(\delta ,z, N)=
\left<G_R(\delta, z, N)\right>_{R,N}.
\label{14o}
\end{equation}
Here, $P_N$ is the known multiplicity distribution for full phase space.

Thus, within the framework of the considered model, we have found 
the multiplicity distribution
in restricted bins from the following global observables
containing dynamical information on the fluctuations:

\medskip
1) The multiplicity distribution $P_R^{\mathrm{conf}}$ for 
the cluster configurations $R$.

2) The multiplicity distribution $P_N$ for particles in the  full
phase space $\Delta$.

\medskip

Of course, both quantities have to be 
defined on the basis of physical knowledge on the dynamical structure of
the particle clustering.

According to (\ref{3o}) and (\ref{3oo}), the
NFMs for (\ref{14o}) have the following form
\begin{equation}
F_q(\delta )=\frac{\left<\tilde F_q(\delta )\right>_{R,N}}
{\left<\tilde F_1(\delta )\right>_{R,N}^q}.
\label{200o}
\end{equation} 
The quantities inside the angular brackets define (usual)
factorial moments for a fixed cluster configuration $R$ and
a fixed total number of particles in the full phase space:
\begin{equation}
\tilde F_q(\delta )=\sum_{q_1=0}^{q}\sum_{q_2=0}^{q_1}\ldots
\sum_{q_{N-1}=0}^{q_{N-2}}C_{q}^{q_1}C_{q_1}^{q_2}
\ldots C_{q_{N-2}}^{q_{N-1}}
\tilde F_{q-q_1}^{\{1\}}\tilde F_{q_1-q_2}^{\{2\}}\ldots
\tilde F_{q_{N-2}-q_{N-1}}^{\{N -1\}}
\tilde F_{q_{N-1}}^{\{N\}}.
\label{19o}
\end{equation}
Here, we have introduced the following definitions of the moments
obtained from $G^{\{m\}}_{N_m}(\delta ,z)$:
\begin{equation}
\tilde F_q^{\{1\}}=N^{[q]}_1p^q ,
\label{20o}
\end{equation}
\begin{equation}
\tilde F_q^{\{m\}}=\sum^{N_m}_{i=q}(m\>i)^{[q]}C_{N_m}^ip^ig^{N_m-i} .
\qquad m > 1.
\label{21o}
\end{equation}

It is convenient to  introduce, instead of $N_m$,
the probability $W_m$ of any particle (chosen at random) to
belong to an $m$-particle cluster. This probability can
be written as
\begin{equation}
W_m=m\,N_m/N ,  \qquad \sum^N_{m=1}W_m(N)=1 .
\label{115o}
\end{equation} 
The set of $W_m$ for $m=1,\ldots ,N$ 
forms the probability distribution which
we shall call the cluster configuration distribution. Hence, $R$
used above represents a particular form of the  
cluster configuration distribution $W_m$.

In fact, the quantities $W_m$ are identical to those used in the theory
of percolation clusters (see \cite{perc} for review). 
There, the cluster is defined as
a group of occupied lattice sites connected 
by nearest-neighbor distances.
Then $N_m/N$ can be treated as the average number (per lattice site)
of $m$-particle clusters and $W_m$ defined from (\ref{115o}) 
is the probability of
any lattice site to belong to a $m$-particle cluster.

\section{Examples}
\label{sec:exm}

Though the theoretical attempt to describe a sample in terms of
discrete distributions is still in a very early stage, one may
already interested in its possible observable consequences.

To illustrate the intermittent properties of the model, let us consider
a specific system in which the particle multiplicity $N$ is fixed 
for all events and 
the $P_R^{\mathrm{conf}}$ has a sharp 
maximum near  $R\simeq\bar{R}$, i.e.,
\begin{equation}
P_R^{\mathrm{conf}}\simeq \mathrm{\delta}_{R,\bar{R}}=
\left\{ \begin{array}{ll} 1, & R=\bar{R},  \\
0, & R\ne\bar{R} . 
\end{array}
\right.
\label{126o}
\end{equation}
Then,
\begin{equation}
\bar G(\delta, z, N)\simeq\prod_{m=1}^{N}G_{\bar{N}_m}^{\{m\}}(\delta ,z).
\label{18o}
\end{equation}
where $\bar{N}_m$ is the most probable cluster configuration.
For a given $N$, expression (\ref{18o})  permits 
us to calculate the NFMs if we know the form
of $\bar{N}_m$ (or the corresponding $\bar{W}_m$).
Then, from (\ref{18o}), we have
\begin{equation}
F_q(\delta )=\frac{\tilde F_q(\delta )}
{\tilde F_1^{q}(\delta )} ,
\label{188o}
\end{equation}
where $\tilde F_q(\delta )$ is given by (\ref{19o}), if
we substitute $\bar{N}_m$ instead of $N_m$.
Note that (\ref{188o}) does not depend on the global particle density $N/\Delta$,
since, for any value of $N$, 
all dependence of the NFMs on the bin size is determined by the 
ratio $\Delta/\delta =M^D$, but not by $\Delta$ itself.

To be specific, we present the results of our calculation of the NFMs for
a few  particular cases:

\medskip
{\em 1) Contribution from two-particle clusters}
\medskip

To study this case, we  use the following simple cluster configuration
\begin{equation}
\bar{W}_1=0.4,  \qquad \bar{W}_2=0.6.
\label{120o}
\end{equation}
In Fig.~\ref{fig1},
we present the results of our calculation for $F_2$ to $F_5$ with
$N=50$ and $D=1$ (closed symbols).
For this value of $N$, according to (\ref{115o}), 
configuration (\ref{120o}) approximately corresponds to
$N_1=20$, $N_2=15$.

From Fig.~\ref{fig1}  we can see that the NFMs have a 
tendency to grow linearly with increasing
of the number of bins $M$. The increase can be fitted 
by the power-law (\ref{1o}) with a  intermittency exponent $\phi_q$ and
a coefficient of proportionality  $a_q$,
\begin{equation}
\ln F_q =\phi_{q}\ln M + \ln a_q , \qquad \phi_{q}>0.
\label{211o}
\end{equation}
Since the main aim  of this section is only to illustrate a real
possibility of the intermittent rise of the NFMs with increasing $M$, we do
not pursue the purpose to fit the results obtained  by (\ref{211o}) for this 
(hypothetical) cluster configuration.

\medskip
{\em 2) Contribution from three-particle clusters}
\medskip

The open symbols in Fig.~\ref{fig1}
represent the behavior of $F_2$ to $F_5$  for 
a sample with three-particle clusters (D=1).
Here,  
\begin{equation}
\bar{W}_1=0.4 , \qquad \bar{W}_3=0.6.
\label{121o}
\end{equation}
This corresponds to $N_1=20$, $N_3=10$ (for $N=50$).
The sample with three-particle clusters has
stronger  intermittent behavior than that obtained using (\ref{120o}).

\section{Bunching-Parameter Method} 

In this section, we argue and numerically show that
the bunching parameters \cite{bp1,bp2}  are more sensitive to the
features of a cluster configuration than the NFMs discussed above.
In fact, the  result of this subsection
complements the statement made in \cite{bp2},  
where it has been  shown that the NFMs are
insensitive to the structure of fluctuations, i.e.
rather different behaviors of the underlying local
multiplicity distribution $P_n(\delta )$  with decreasing
$\delta$ can have the same  intermittent trend of the
NFMs. 

The definition of the 
bunching parameters $\eta_q$ is \cite{bp1,bp2} 
\begin{equation}
\eta_q=\frac{q}{q-1}\frac{P_q(\delta )P_{q-2}(\delta )}
{P_{q-1}^2(\delta )},
\label{bp11}
\end{equation}
where $P_n(\delta )$ is the probability of having $n$ particles
inside a restricted phase-space bin $\delta$. 
For example, for a Poisson distribution,  one has
$\eta_q(\delta )=F_q(\delta )=1$, so that $q$th order BP
measures  the  deviation of the local multiplicity
distribution from the Poisson.
In general, independent particle production 
leads  to $\eta_q(\delta )\ne 1$ for $\delta\to 0$. 
In this case, however,
the bunching parameters are $\delta$-independent constants.
For  example, 
if all spikes  are purely statistical
for events with a fixed finite  multiplicity $N$,
i.e., when the cluster
configuration reduces to the trivial case
\begin{equation}
\bar{W}_m=\mathrm{\delta}_{m,1},
\label{2rto}
\end{equation}
then (\ref{11o}) becomes  a positive binomial distribution  and
the corresponding BPs have the following $\delta$-independent form
\begin{equation} 
\eta_q=\frac{q - 1 - N}{q - 2 - N}.
\label{bp12}
\end{equation}

The interest in the use of BPs to extract  information on the cluster
configuration lies in the fact that the BP of rank $q$ is sensitive only
to multiparticle clusters with $m\le q$ particles. That is, a given  BP
acts as a filter for clusters having a small number of particles.
This follows directly from the definition (\ref{bp11}).
On the contrary, the NFM of rank $q$ is sensitive to spikes
with $n\ge q$ particles and acts as a filter for clusters
with large number of particles. 
This means that the bunching parameters have a complementary 
property which is very important to obtain a 
refined insight into various
multiparticle systems with intermittent behavior.

To  demonstrate this point,
we shall use 
(\ref{18o})
and the  relation $P_q(\delta )=G^{(q)}(z)\mid_{z=0}/q!$.
Note that for this calculation we can use the expression
(\ref{19o}),  substituting $P_q(m,N_m)$ (see (\ref{100o}))
multiplied by the
factor $q!$, instead of the $\tilde F_q^{\{m\}}$.

In  Fig.~\ref{fcom1} (a),(b),  we present the behavior of the NFMs 
for $\bar{W}_1=0.4$, $\bar{W}_3=0.6$ (only monomers and three-particle clusters
exist) and $\bar{W}_1=0.22$, $\bar{W}_2=0.4$, 
$\bar{W}_3=0.3$, $\bar{W}_4=0.08$.
The latter configuration represents a sample with
two-, tree- and  four-particle clusters. The behavior of
the NFMs is the same for  these rather different samples, not
only qualitatively, but also quantitatively (actually,
these configurations were chosen for an illustrative purpose, 
because  they exhibit 
the same behavior of the NFMs). 
In contrast to the NFMs, 
the behavior of the BPs
is   different for the same  cluster configurations
(see Fig.~\ref{fcom2} (a) (b)).  
This means that the bunching parameters provide 
a simple and effective method to analyze models and compare
them with experimental data. 

This point  has already been illustrated in 
\cite{bp2} using a Monte-Carlo model,
where we have shown that systems  with a similar  trend of
the NFMs exhibit  different behavior when one studies them with
the help of bunching parameters. 

\section{Conclusion}
\label{sec:con}

We have presented a statistical analysis of  multiparticle
fluctuations  in
ever smaller  phase-space bins for different dimensions using
the characteristics of primarily observable - multiparticle clusters.
This new theoretical method
links the {\em global} characteristics
(a cluster configuration distribution and a multiplicity
distribution of particles in the full phase space) of the sample 
to {\em local} 
multiplicity fluctuations of a multiparticle process.
As we have seen,
the theoretical approach developed in its simplest form
(no correlations between clusters)
reveals many promising features.
The model can lead to an intermittent rise
of the NFMs with decreasing $\delta$,
even for the simplest special cases of the global 
characteristics 
taken for illustrative purposes.
In our model, we  ``a priori'' take into account the clusters,  
without any physical examination of the 
clustering phenomenon itself, which,
up to now, is
a standard topic in soft hadron physics where the application
of QCD is very restricted.

One of the reasons for the approach presented in this paper is to
give an exact method to derive
the intermittent behavior in various cluster models.
In particular,  
the general method developed here 
can  be used to study intermittent behavior of 
the statistical models \cite{onm} 
of a fragmentation process 
in usual phase-space
variables, rather then for mass distributions. 

There is an interesting point connected with our approach. In fact,
our prescription is to 
decompose the local fluctuations into clusters with 
varying number of particles. Apparently this 
situation is closely analogous to that in which
the intermittent signal is decomposed into contributions from clusters living on
specific scales (the so-called method of orthogonal wavelet transform used
in high-energy physics \cite{gr1}). Note, however, that the
connection between that method and our approach is not trivial, 
since the conceptual distinction between
these strategies is mathematically clear-cut. 

\medskip

Acknowledgments

This work is part of the research program of the ``Stichting voor
Fundamenteel Onderzoek der Materie (FOM)'', which
is financially supported by the ``Nederlandse Organisatie voor
Wetenschappelijk Onderzoek (NWO)''.
We are indebted to Prof. W.Kittel and Prof. J.-L.Meunier 
for suggesting improvements. 
The research 
of V.I.K. was supported by Grant F95-023 of the 
Fund for Fundamental Research of the Republic of Belarus.

{}

\newpage
\begin{center}

\vspace{2.0cm}

\begin{figure}
\begin{center}
\mbox{\epsfig{file=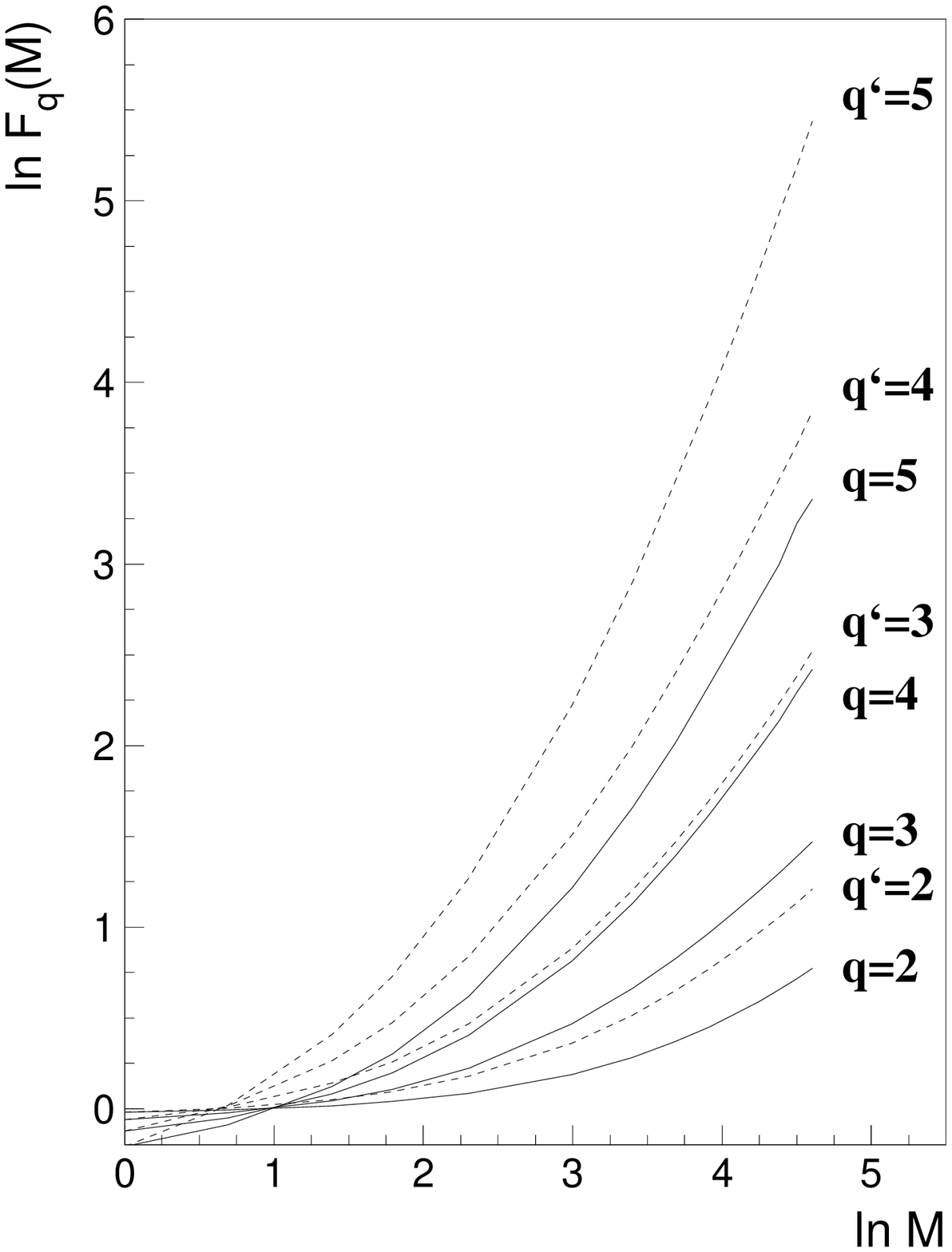,width=9.0cm}}
\caption[fig1]
{\it
NFMs as a function of the number of bins $M$.
Solid lines represent the configuration $\bar{W}_1=0.4$, $\bar{W}_2=0.6$,
dashed lines correspond to $\bar{W}_1=0.4$, $\bar{W}_3=0.6$.
For both cases $N=50$.
}
\label{fig1}
\end{center}
\end{figure}

\newpage

\begin{figure}
\begin{center}
\mbox{\epsfig{file=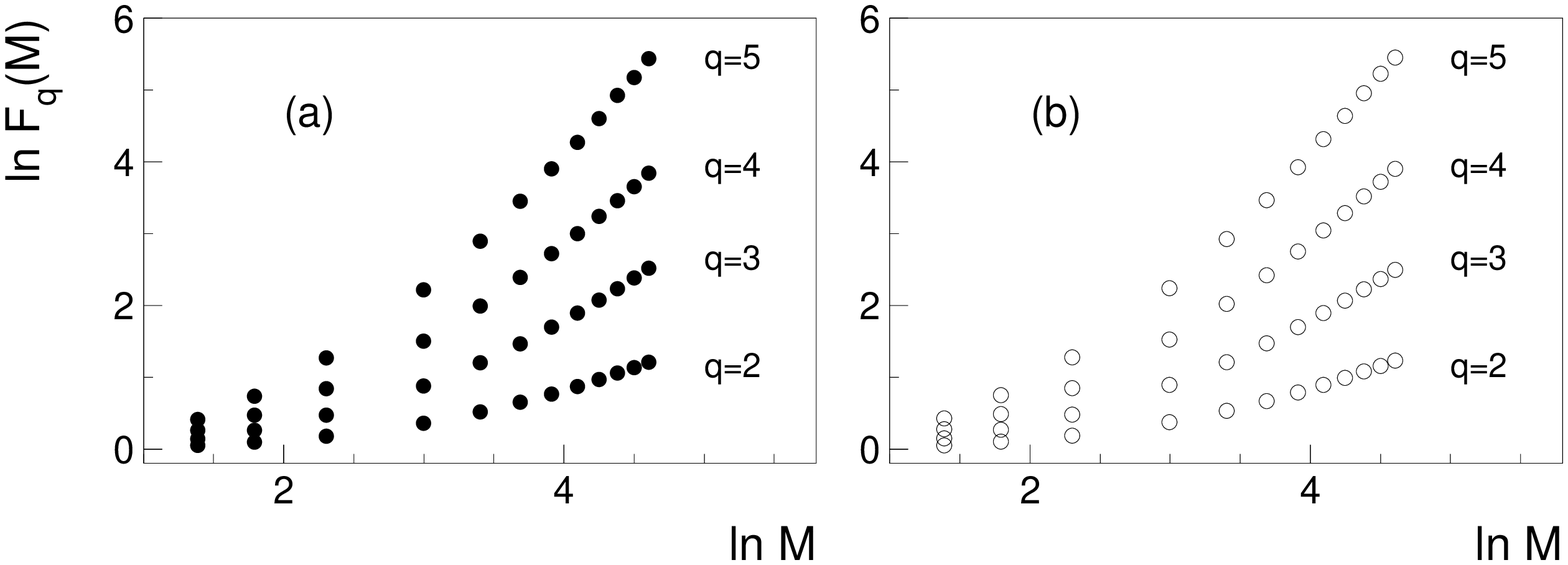,height=6.0cm}}
\caption[fcom1]
{\it Dependence of $\ln F_q(M)$ ($q=2,\ldots ,5$) on $\ln M$. 
{\bf (a)}: $\bar{W}_1=0.4$, $\bar{W}_3=0.6$ ;
{\bf (b)}: $\bar{W}_1=0.22$, $\bar{W}_2=0.4$, 
$\bar{W}_3=0.3$, $\bar{W}_4=0.08$.
Here, we set $N=50$ for both cases.}
\label{fcom1}
\mbox{\epsfig{file=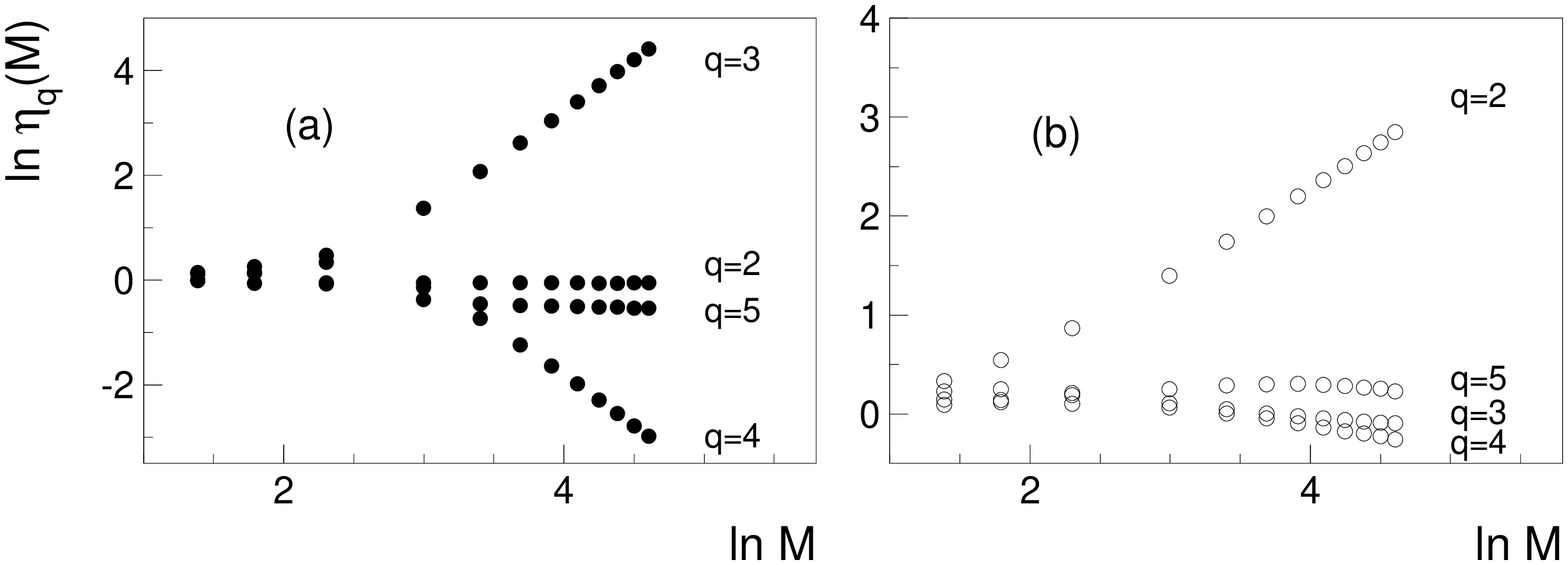,height=6.0cm}}
\caption[fcom2]
{\it Dependence of $\ln\eta_q(M)$ ($q=2,\ldots ,5$) on $\ln M$ for
the same cluster configurations as in Fig.~\ref{fcom1}}.
\label{fcom2}
\end{center}
\end{figure}

\end{center}
\end{document}